\documentclass[manuscript]{aastex62}

\graphicspath{{./}{figures/}}

\submitjournal{ApJ}

\usepackage{footnote}
\usepackage{savesym}
\savesymbol{tablenum}
\usepackage{siunitx}
\restoresymbol{SIX}{tablenum}
\usepackage{float}
\floatstyle{plaintop}
\restylefloat{table}
\usepackage{url}

\begin{document}

\title{First evidence of enhanced recombination in astrophysical environments and the implications for plasma diagnostics}

\correspondingauthor{Ahmad Nemer}
\email{anemer@princeton.edu}

\author[0000-0002-0786-7307]{A. Nemer}
\affil{Auburn University, Auburn, AL}
\affil{Princeton University, Princeton, NJ}

\author{N. C. Sterling}
\affiliation{University of West Georgia, Carrollton, GA}

\author{J. Raymond}
\affiliation{Center for Astrophysics $\mid$ Harvard {\&} Smithsonian, Cambridge, MA}

\author{A.K. Dupree}
\affiliation{Center for Astrophysics $\mid$ Harvard {\&} Smithsonian, Cambridge, MA}

\author{J. Garc\'{\i}a-Rojas}
\affiliation{Instituto de Astrof\'{\i}sica de Canarias, La Laguna, Tenerife, Spain}
\affiliation{Universidad de La Laguna, La Laguna, Tenerife, Spain}

\author{Qianxia Wang}
\affiliation{Auburn University, Auburn, AL}
\affiliation{Rice University, Houston, TX}

\author{M.S. Pindzola}
\affiliation{Auburn University, Auburn, AL}

\author{C. P. Ballance}
\affiliation{Queen's University of Belfast, Belfast, UK.}

\author{S.D. Loch}
\affiliation{Auburn University, Auburn, AL}

%% Note that the \and command from previous versions of AASTeX is now
%% depreciated in this version as it is no longer necessary. AASTeX 
%% automatically takes care of all commas and "and"s between authors names.

%% AASTeX 6.2 has the new \collaboration and \nocollaboration commands to
%% provide the collaboration status of a group of authors. These commands 
%% can be used either before or after the list of corresponding authors. The
%% argument for \collaboration is the collaboration identifier. Authors are
%% encouraged to surround collaboration identifiers with ()s. The 
%% \nocollaboration command takes no argument and exists to indicate that
%% the nearby authors are not part of surrounding collaborations.

%% Mark off the abstract in the ``abstract'' environment. 
\begin{abstract}

We report the first unambiguous observational evidence of Rydberg Enhanced Recombination (RER), a potentially important recombination mechanism that has hitherto been unexplored in low-temperature photoionized plasmas. RER shares similarities to dielectronic recombination, with the difference that the electron is captured into a highly excited state below the ionization threshold — rather than above the threshold — of the recombining ion.  We predict transitions of carbon and oxygen ions that are formed via the RER process, and their relative strengths with collisional-radiative spectral models.   Optical C II RER features are detected in published high-resolution spectra of eight planetary nebulae, and a C III transition has been found in the UV spectrum in a symbiotic star system.  The relative intensities of these lines are consistent with their production by this recombination mechanism.  Because RER has not previously been accounted for in photoionized plasmas, its inclusion in models can significantly impact the predicted ionization balance and hence abundance calculations of important astrophysical species. Calculations for $C^+$ suggest that the enhancement in the total recombination rate  can amount to a factor of 2.2 at 8100 K, increasing to 7.5 at $T_e$=3500 K. These results demonstrate the importance of including RER in models of photoionized astrophysical plasmas and in elemental abundance determinations.

\end{abstract}

%% Keywords should appear after the \end{abstract} command. 
%% See the online documentation for the full list of available subject
%% keywords and the rules for their use.

%% From the front matter, we move on to the body of the paper.
%% Sections are demarcated by \section and \subsection, respectively.
%% Observe the use of the LaTeX \label
%% command after the \subsection to give a symbolic KEY to the
%% subsection for cross-referencing in a \ref command.
%% You can use LaTeX's \ref and \label commands to keep track of
%% cross-references to sections, equations, tables, and figures.
%% That way, if you change the order of any elements, LaTeX will
%% automatically renumber them.
%%
%% We recommend that authors also use the natbib \citep
%% and \citet commands to identify citations.  The citations are
%% tied to the reference list via symbolic KEYs. The KEY corresponds
%% to the KEY in the \bibitem in the reference list below. 

\section{Introduction}

A new recombination process, proposed by \citet{Robicheaux2010}, modifies current descriptions of electron-ion recombination in plasmas to include a process that we refer to as Rydberg Enhanced Recombination (RER). This new mechanism may have far reaching implications, affecting elemental abundance determinations and ionization equilibrium solutions. However, there was no previous evidence for RER in laboratory or astrophysical plasmas. In this paper we present spectroscopic evidence of RER in planetary nebulae and symbiotic stars, and explore its effects on their physical conditions and chemical compositions. \\

The charge state distribution of elements in astrophysical nebulae results from a balance between photoionization and recombination processes. The main recombination processes in photoionized plasmas are non-resonant, radiative recombination and resonant, dielectronic recombination (DR). DR is often the dominant process and involves a free electron being captured into a doubly excited state (resonance), which then decays to a lower level via photon emission. \citet{Robicheaux2010} introduced RER, which includes resonances just below the ionization threshold that are not considered in DR.
RER is analogous to DR, but includes three steps (indicated in Fig. \ref{fig:RER}); 1) low energy free electrons recombine to high-n singly-excited states of an ion (called Rydberg states); 2) radiationless transition to a resonance below the threshold (which we call dielectronic auto-transfer); 3) radiative transitions to lower levels to complete the process.   \\

Photoionized nebulae are ubiquitous, diagnostically important objects \citep{Peimbert2017}, and have contributed significantly to our knowledge of the chemical evolution of the cosmos. These plasmas are ionized by low-mass dying stars in the case of planetary nebulae (PNe), a hot compact object in symbiotic binaries, or young, massive stars in H II regions. Accurate elemental abundance determinations are essential in bench-marking existing models of stellar evolution, nucleosynthesis, galactic composition and kinematics, and cosmology \citep{Zaritsky1994,Hamann1999,Savin2000,Henry2018}. 
However, plasma diagnostics suffer from various issues, including discrepancies in temperature and abundance determinations \citep{Peimbert2017}. \citet{Ferland1998} showed that the lack of reliable DR rates is the dominant uncertainty in ionization balance calculations of photoionized plasmas. Additionally, ionization correction factors are used to convert ionic abundances into elemental abundances \citep{Delgado-Inglada2014}. Such corrections for unobserved ions depend on accurate ionization and recombination rate coefficients. Neglecting the contribution of RER to recombination thus leads to inaccurate abundance determinations. This will affect abundance ratios such as those used to constrain the intergalactic medium \citep{Savin2000} and nucleosynthesis models \citep{Henry2018}. \\ 

The physical conditions of low temperature ($\sim$10,000 K) photoionized nebulae make them ideal `astrophysical laboratories' to search for emission lines produced by RER, given the low electron temperatures needed for RER. In this letter, we identify ions that could be affected by RER, and present spectroscopic evidence of RER using high resolution optical spectra of PNe and UV spectra of a symbiotic binary. \\

\section{Background theory} \label{sec:theory}

\begin{figure}[!hb]
\centering
\frame{\includegraphics[scale = 0.525]{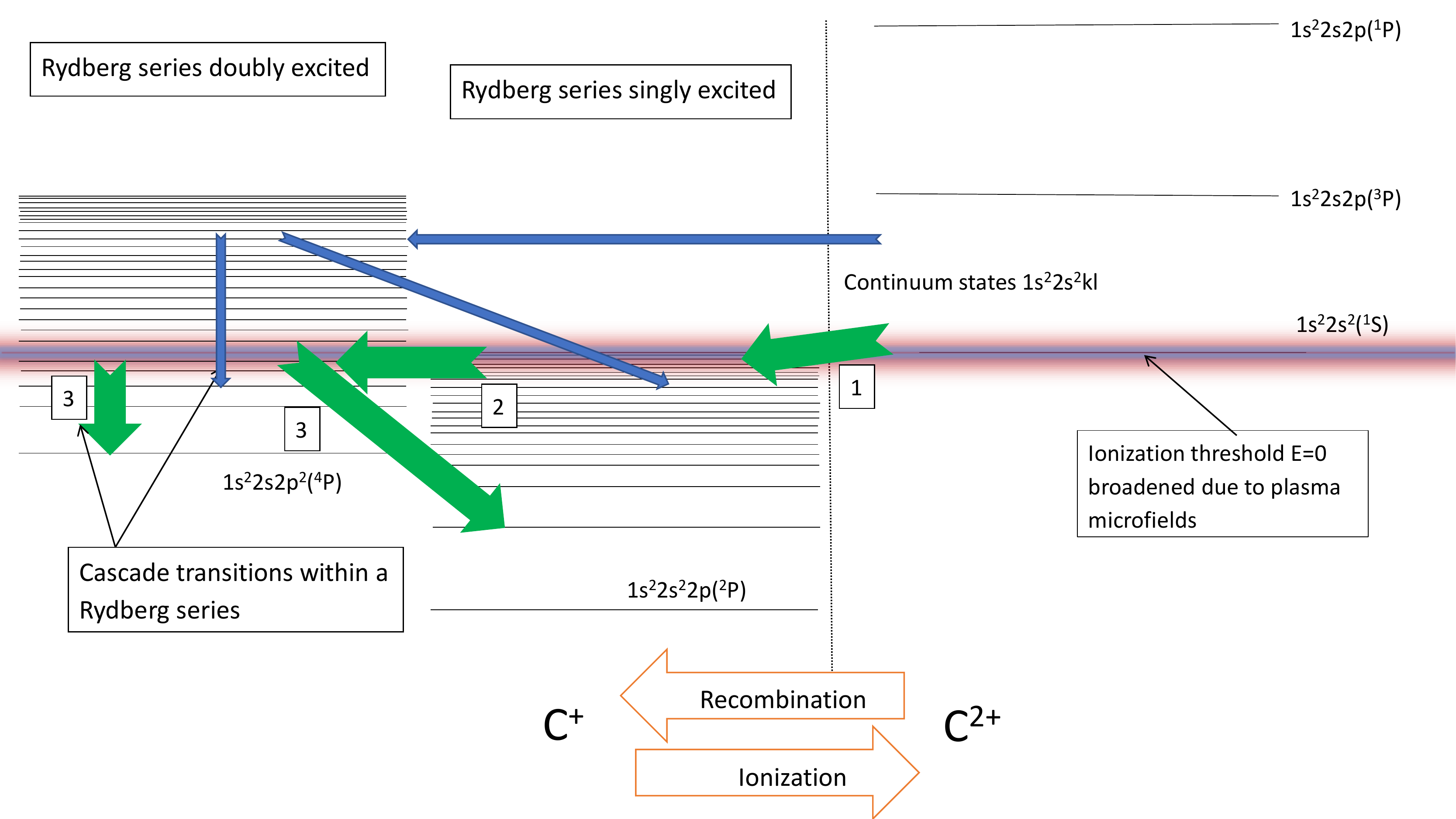}}
\caption{Energy level diagram. The right hand-side shows C$^{2+}$ recombining into C$^+$ (left side). The blurred line represents the ionization threshold (C$^{2+}$ ground state). Two series of energy levels are shown, the singly excited $1s^22s^2nl$ series, and the doubly excited $1s^22s2pnl$ series. The thin blue arrows trace DR, and the thick green arrows denote RER, with numbered steps defined in the text.
\label{fig:RER}}
\end{figure}

Fig. \ref{fig:RER} illustrates the DR and RER processes. Historically, DR has been assumed to be initiated only from free electron states in the continuum (blue arrows in Fig. \ref{fig:RER}). \citet{Robicheaux2010} indicated that a similar process, RER, is possible between Rydberg states and below-threshold resonances of the same ion (Fig. \ref{fig:RER}, green arrows). Such transitions have been observed in charge exchange experiments \citep{Ali2016}. Quantum mechanically this is a result of wavefunction mixing, with the ion in a superposition of each of the states. 
Because RER involves low-energy electron capture into Rydberg states, this process is most effective in low temperature plasmas. \\

To test for the occurrence of RER in astrophysical nebulae, it is necessary to predict spectral lines produced by this mechanism. Quantifying the effect of RER on the ionization equilibrium additionally requires the reevaluation of total recombination rate coefficients. To address these needs, two multi-configuration atomic structure codes were modified, the Breit-Pauli code AUTOSTRUCTURE \citep{Badnell2011} and the Dirac-Fock code GRASP \citep{Dyall1989}. \\

As shown by \citet{Robicheaux2010}, calculating RER rates is analogous to the DR methodology. DR cross sections are converted to plasma rate coefficients via integration over a Maxwellian distribution of free electrons. In contrast, RER involves Rydberg states, and the rate coefficients are computed by integrating over their population distributions. The capture rate from the Rydberg states into a resonance ($j$) below the ionization threshold is
\begin{equation}
    \alpha_j^{below}(T_e) = \left(\frac{4 \pi a_o^2 I_H}{k_B T_e}\right)^{3/2} \frac{\omega_j}{\omega_+} e^{-E_c/k_BT_e} A^a_{j,E_c} b_j
\end{equation}
\noindent where $I_H$ is the Rydberg constant, $k_B$ is Boltzmann's constant, $T_e$ is the plasma electron temperature, $\omega_j$ and $\omega_+$ are the statistical weights of the doubly excited state and  ground state of the recombining ion, respectively, $E_c$ is the energy of the resonance relative to the ionization potential, $A^a$ is the dielectronic auto-transfer rate to Rydberg states and $b_j$ is the departure factor (described below). To obtain the enhancement to the total recombination rate for that charge state, the rate coefficient is multiplied by a radiative branching ratio and summed over all below threshold states $j$ that can mix with the Rydberg states.\\

Resonances below the ionization potential are therefore populated by RER (and other processes discussed later), and depopulated by radiative decay ($A^r$) and dielectronic auto-transfer to Rydberg states ($A_a$). Thus the population of the doubly excited state is
\begin{equation}
N_j = \frac{N^+ n_e \alpha_j^{RER} + N^+ n_e \alpha_j^{cascade} + N^g n_e q_j^{fluorescence}}{\sum_k A_r + \sum_l A_a},
\end{equation}
where $N^+$ and $N^g$ are the populations of the ground states of the recombining and recombined ions, and $\alpha_j$ and $q_j$ represent the effective rates to populate the resonance. This population density is multiplied by spontaneous emission rates to evaluate the emission per unit volume per second for the below threshold transitions. \\

We utilized collisional-radiative theory, with a modified version of the ADAS code \citep{Giunta2012}, to determine the Rydberg populations. ADAS204 was used as it includes all collisional and radiative populating processes for the Rydberg states (dominated by three-body and radiative recombination) and depopulating processes (mainly collisional and photoionization), then solves the quasi-static equilibrium equations to produce level populations. Rydberg populations are described in terms of a departure coefficient, defined as the ratio between the level populations and their local thermal equilibrium (LTE) value. The highest n-shells are close to LTE, and below about n=200 the populations fall below the LTE value due to less frequent collisions and faster radiative decay. \\

A spectral synthesis code was developed that uses these atomic data to model the intensities of the RER spectral lines, and accurate energies \citep[NIST database][]{Reader2012} to predict the wavelengths. 

\section{Results} \label{sec:Results}

\subsection{Predicted RER emission lines}

We established criteria to identify transitions whose upper levels are likely populated by RER and thus may produce detectable emission lines. The doubly-excited upper levels must i) be propinquitous to the ionization threshold; ii) mix with Rydberg states; and iii) be from an ion widely observed in astrophysical nebulae. We investigated the first two rows of the periodic table for transitions that fit these criteria, and C II, C III, O II, and O III were the most promising, with accurate energy levels available for the resonances. Table \ref{tab:lines} shows the candidates for the search.

\begin{table}[htb]
\begin{center}
\begin{tabular}{|c|c|c|c|}
\hline
    & Term (multiplet) & Energy below                & Wavelengths of  \\
Ion &                    & ionization potential (eV) & strongest transitions \footnote{Wavelengths shorter than 2000 \si{\angstrom} are in vacuum,  and wavelengths above 2000 are in air. Observed lines are highlighted in bold.} (\si{\angstrom}) \\
\hline
C II & 1s$^{2}$2s2p($^{3}$P$^{°}$)3d ($^{4}$D) & 0.010097-0.011355 & 651.21, 651.39\\
\hline
C II & 1s$^{2}$2s2p($^{3}$P$^{°}$)3d ($^{4}$F) 
& 0.105355-0.112927 & \textbf{7115.53}, 7134.03,\\
 &   &  & \textbf{7112.94}, \textbf{7119.73}\\
\hline
C III & 1s$^{2}$2p4p ($^{1}$D) & 0.07475-0.05254 & 1581.43, 416.77,\\
      &                    &                 & 1623.25, 1512.93\\
\hline
C III & 1s$^{2}$2p4p ($^{3}$F) & 0.05148 & 1593.65, 1594.27,\\
      &                    &         & 2799.19, \textbf{1553.38}\\
\hline  
O II & 1s$^{2}$2s$^{2}$2p$^{2}$($^{1}$S)4s ($^{2}$S) & 0.130312 & 413.65 \\
\hline
O II & 1s$^{2}$2s$^{2}$2p$^{2}$($^{1}$D)5s ($^{2}$D) & 0.635618 & 397.9, 420.7\\
\hline
O III & 1s$^{2}$2s$^{2}$2p$^{2}$($^{4}$P)4p ($^{3}$D) & 0.04554 & 225.9, 671.8\\
\hline
O III & 1s$^{2}$2s$^{2}$2p$^{2}$($^{4}$P)4p ($^{5}$P) & 0.45183-0.59911 & 2430.3, 2427.66, 2426.78\\
\hline
\end{tabular}
\end{center}
\caption{C and O transitions arising from RER.} 
\label{tab:lines}
\end{table}

Energies from the NIST database were used for the wavelength predictions. We focused on the C II lines $\lambda$7112.94, 7119.73 and 7115.53 due to the availability of high-resolution optical spectra of PNe, and C III $\lambda$1553.38 using high-resolution UV spectra of symbiotic binaries. The UV RER lines are generally predicted to be stronger than optical lines due to higher dielectronic auto-transfer rates and branching ratios. 

\subsection{Spectroscopic Evidence for RER}

High-resolution spectroscopy is critical for identifying RER lines, both to resolve these lines from nearby features of other species and to provide greater line-to-continuum contrast for weak features. We searched for optical C II RER lines (Table \ref{tab:lines}) in published high-resolution spectra of seven PNe observed with 4-8 m telescopes: IC 418 \citep{Sharpee2003}; IC 2501, IC 4191, NGC 2440, NGC 7027 \citep{Sharpee2006}; NGC 6369 \citep{Garcia-Rojas2012}; and NGC 3918 \citep{Garcia-Rojas2015}.  We also studied an unpublished spectrum of Hb 12, obtained with the 2D coud\'{e} spectrograph on the 2.7-m Harlan J. Smith Telescope at McDonald Observatory \citep{Sherrard2017}. We show six of these objects that exhibit C II RER lines (Fig. 2a), whose intensities relative to H$\beta$ are given in Table \ref{tab:lines2}. \\

\begin{table}[b]
\begin{center}
\begin{tabular}{|c|c|c|c|c|c|c|}
\hline
PN	&10$^{-3}I$(7115)/$I$(H$_{\beta}$)	& 10$^{-3}I$(3876)/$I$(H$_{\beta}$)   & 10$^{-3}I$(5259)/$I$(H$_{\beta}$) &     RER  & $I$(7112)/ \\ 
    &  	                                   &                        &              	& Contribution     &  $I$(7115) \footnote{The last column shows the observed ratio of the strongest C II RER multiplet lines.} \\
    &  	                                   &                        &              	& to $I$(7115)  \footnote{See text for discussion of negative numbers}     &   \\
\hline
Hb 12	&4.48 $\pm$ \num{0.90}  &	\nodata	&\nodata	&1.00& 1.76 $\pm$ \num{0.50}\\
\hline
IC2501	&9.80 $\pm$ \num{2.94}& 14.00 $\pm$ \num{1.50}	&5.00 $\pm$ \num{1.00} 	    &0.22& 0.76 $\pm$ \num{0.32}\\
\hline
IC418	&4.30 $\pm$ \num{0.43}&	6.90 $\pm$ \num{1.38}	&6.33 $\pm$ \num{1.26}	&-0.34& 1.21 $\pm$ \num{0.17}\\
\hline
IC4191	&3.10 $\pm$ \num{0.62}  & 96.00 $\pm$ \num{10.10}	&\nodata	&-10.00& 0.90 $\pm$ \num{0.25}\\
\hline
NGC2440	&23.00 $\pm$ \num{4.60} &	\nodata	&3.60 $\pm$ \num{0.72} &0.92&1.13  $\pm$ \num{0.32}\\
\hline
NGC3918	&6.80 $\pm$ \num{1.36}  & \nodata	&\nodata	&1.00& 1.04 $\pm$ \num{0.29}\\
\hline
NGC6369	&11.00 $\pm$ \num{4.40} &	\nodata	&\nodata	&1.00& 0.636 $\pm$ \num{0.36}\\
\hline
NGC7027	&14.70 $\pm$ \num{1.47} &	\nodata	&7.90 $\pm$ \num{1.58} &0.72& 1.22 $\pm$ \num{0.17}\\
\hline
\end{tabular}
\end{center}
\caption{Intensities (on the scale $I$(H$_{\beta}=100$)) of the C II RER line $\lambda$7115.53 and cascade lines $\lambda$3876 and 5259 that compete with RER to populate these levels.}
\label{tab:lines2}
\end{table}

To verify the identities of C II $\lambda\lambda$7112.94, 7119.73 and 7115.53, we utilize the Atomic Line List v2.05b21 \footnote{\url{http://www.pa.uky.edu/~peter/newpage/}} to search for other possible identifications with rest wavelengths within 1 \si{\angstrom}. We considered forbidden transitions of atomic ions with excitation energies below 10 eV, and permitted transitions of elements in the first three rows of the periodic table.  For potential alternative identifications, we searched for multiplet members, lines from the same upper level, and for forbidden transitions of iron-peak elements, the strongest optical lines expected given the physical conditions of the nebulae. Using this method, we ruled out alternative identifications for these C II lines, in agreement with the identifications of \citet{Storey2013}. Based on our vetting criteria and the detection of multiplet members, we are confident that C II $\lambda\lambda$7112.94, 7119.73 and 7115.53 have been detected in each of the eight PNe we investigated.\\

It is also important to confirm that the C II RER lines are not produced by another mechanism. Fluorescence is quantum mechanically forbidden to populate these resonances from the ground (or metastable) states. Alternatively, cascade from higher (i.e. above threshold, populated by DR) doubly-excited states in the same Rydberg series (e.g. $1s^22s2pnl$) can provide a populating mechanism. We searched the spectra for cascade lines that can populate the upper level of C II $\lambda$7115.53 based on their branching ratios, namely C II 3876 and 5259 \si{\angstrom} (Table \ref{tab:lines2}). In five objects (NGC 6369, NGC 7027, NGC 2440, NGC 3918 and Hb-12), the cascade lines were either not detected, or are too weak to explain the observed intensities of the RER lines. This indicates that the C II multiplet $\lambda\lambda$7112.94, 7115.53, and 7119.73 is populated almost entirely by RER. In two PNe (IC 418 and IC 2501), the cascade contribution was close to the observed intensity in the two lines, indicating that RER has a negligible contribution. In IC 4191, the cascade line intensities are much larger than predicted compared to that of $\lambda$7115.53, suggesting that C II $\lambda\lambda$3876 and 5259 are either blended with unknown features or misidentified. When the summed cascade portion of the C II $\lambda$7115.53 intensity is larger than that observed, we give a negative factor in Table \ref{tab:lines2}. \\

For a quantitative comparison of observed and predicted line intensities, it is necessary to consider both observational and theoretical uncertainties. The emission lines produced by RER are weak, and the error bars on their observed intensities are $20-25\%$ in the objects investigated. Moreover, flux calibration of cross-dispersed echelle spectra can lead to additional systematic uncertainties. From the atomic physics perspective, the dielectronic auto-transfer rates are uncertain by $\sim 30\%$, depending on the coupling model adopted to compute them. The largest uncertainty in our predicted RER line intensities and modified recombination rate coefficients is likely in the computed Rydberg populations. These were calculated from collisional-radiative models that require accurate knowledge of the plasma conditions (e.g., temperature and density) of the emitting region(s), and 
comprehensive inclusion of RER. Our model predictions assumed uniform electron temperature and density, thus neglecting local variations of these parameters \citep{Liu2000}. \\

In Table \ref{tab:lines2} we show the observed intensity ratios of C II 7112.94 and 7115.53 \si{\angstrom}, which range between 0.76 and 1.78 (with an average of 1.15). These values agree with the predicted ratio of 0.63 to within the observational and modeling uncertainties estimated to be 40\%. \\

\begin{figure}[ht!]
\hspace*{0cm}
\includegraphics[scale = 0.85]{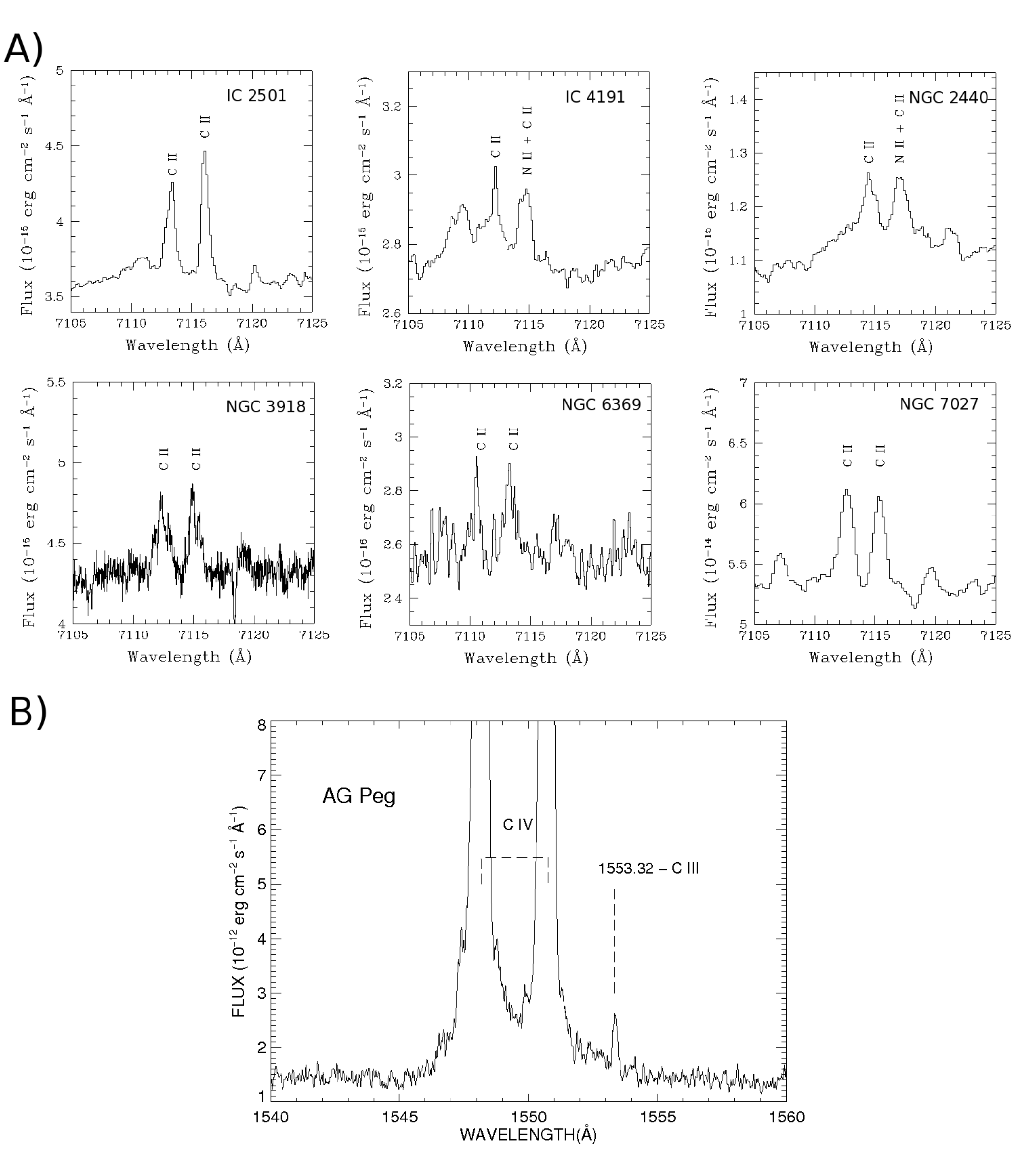}
\caption{A: High-resolution spectra showing the C II $\lambda\lambda$7112.94 and 7115.53 RER lines in the PNe IC 2501, IC 4191, NGC 2440, NGC 7027, NGC 3918, and NGC 6369. B: C III $\lambda$1553.32 in the \textit{HST/GHRS} z27e020at spectrum (program 5360, principal investigator Nussbaumer) of AG Pegasi.
\label{fig:spectra}}
\end{figure}

Because RER transitions in the UV are predicted to be stronger than their optical counterparts, we investigated archival UV spectra of PNe and symbiotic binaries. Of the PNe whose optical spectra we investigated, sufficiently high-resolution \textit{IUE} spectra exist for Hb 12, IC 418, NGC 3918, and NGC 7027.  The individual datasets do not exhibit RER lines, but the low S/N of the spectra do not provide strong constraints. \\

Symbiotic stars are binary systems in which a white dwarf accretes gas from a red giant. We analyzed the UV spectrum of the AG Pegasi symbiotic, whose luminosity decreased by a factor of four between 1984 and 1994 \citep{Eriksson2006}, as the broad C IV resonance lines from the white dwarf wind declined. The C III $\lambda$1553.8 RER line was not visible in the earlier spectrum, but became apparent after the C IV emission faded (Fig. 2b). The\textit{ HST}-GHRS observations used by \citet{Eriksson2006} clearly revealed the 1553~\si{\angstrom} line, but it was not identified by those authors. We verified the identity of this feature in a manner similar to that for the optical C II lines, comparing against the line lists in \citet{Eriksson2006}, and visual inspection of the IUE SWP47715 spectrum (program PA047, principal investigator Vogel). Due to the preponderance of fluorescently-excited Fe and Co transitions in the UV spectrum of AG~Peg, we expanded our search to include permitted lines from elements up to Zn, but found no plausible alternate identifications. \\

In summary, we identify C II multiplet $\lambda\lambda$7112.94, 7115.53, and 7119.73 lines in eight PNe, and show that RER is the dominant populating mechanism for the upper levels of these transitions in five objects.  The RER line C III 1553.38 \si{\angstrom} is seen in the AG Pegasi symbiotic binary. These observations and our analysis represent the first empirical evidence of the RER mechanism. \\

\section{Implications of RER for photoionized plasmas}

\subsection{Effects of RER on Abundance Determinations and the Ionization Balance of Nebulae} 

The inclusion of RER in the analysis of astrophysical plasmas is expected to change the ionization equilibrium, due to an increased rate of recombination, and hence elemental abundances. \\

\begin{figure}[ht!]
\hspace*{-0.7cm}
\includegraphics[scale = 0.58]{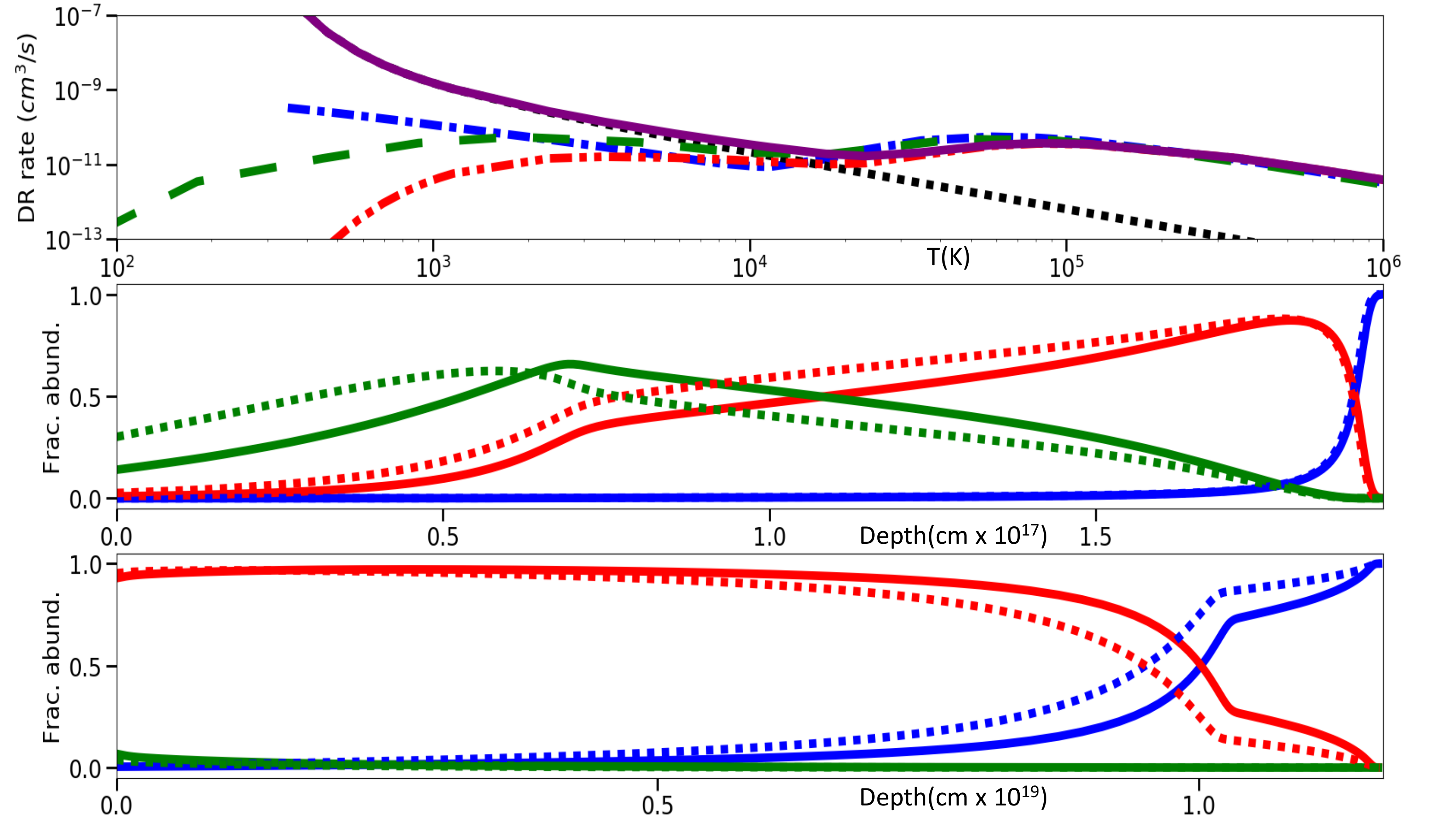}
\caption{Top: Comparison amongst C$^{2+}$ DR rates from the literature. The blue (dash-dotted) curve represents Cloudy rates. The green (dashed) curve is our calculated DR rates, the red (dash-dot-dot) curve is measured DR rates, and the black (dotted) curve is our calculated RER rates. The purple (solid) curve is the recommended total rates. Note the substantial enhancement at T$<$10$^4$K which will decrease the fraction of C$^{2+}$ ions. Middle: Fractional abundances of carbon ions as a function of depth in the PN test suite (dotted with old rates and solid with new rates):  the blue curve is C$^{+}$, red is C$^{2+}$ and the green curve is C$^{3+}$. Bottom: Same as middle, for the H II region Paris model.
\label{fig:cloudy}}
\end{figure}

To demonstrate these effects, we modified the DR rate coefficients in Cloudy \citep{Ferland2017} to include contributions from RER. Fig. \ref{fig:cloudy} illustrates the modified recombination rate coefficients for C$^{2+}$, comparing those used in Cloudy \citep{Colgan2003} to our calculations and to measured rates \citep{Fogle2005}; the recommended rates are a hybrid of measured and calculated values. Below $\sim$10,000K recombination rates are dominated by below-threshold resonances, and RER produces rates that are substantially higher than other recombination processes. The enhancement to the recombination rates due to RER can be seen from the difference between the measured rates (red) and the recommended rates (purple), a factor of 2.2 at T=8100 K, and 7.5 at T=3500K. \\

In our models, RER only modifies the populations of the upper levels of transitions in Table \ref{tab:lines} (and hence their emissivities), but RER should be included in a comprehensive collisional radiative modelling to predict its effects on the populations of all excited states (and hence their emission spectra). C$^{2+}$ abundances are often deduced from optical lines like C II $\lambda$4267, for which the effect of RER is yet to be determined. To illustrate how RER can affect line ratios (and hence abundance determinations), we calculated the line ratio of C II $\lambda$4267 to C II $\lambda$7115 with and without RER as a function of temperature. We found that the ratio is reduced by two orders of magnitude at T=5000K, and by an order of magnitude at T=10,000K due to the enhancement of C II $\lambda$7115 line emissivity. The effect on other C II lines requires full inclusion of RER in emissivity calculations, and incorporating density and temperature inhomogenieties \citep{Liu2000} in radiative transfer models.

We evaluated models for the “Paris Meeting” PN, H II region, and AGN narrow line region simulations in the Cloudy test suite\footnote{see www.nublado.org}, both with and without the modified recombination rate coefficients for C$^{2+}$ and C$^{3+}$.  In the PN model, we introduced density fluctuations spanning $10^3-—10^4 cm^{-3}$. There is a notable shift in the fractional abundance curves of C$^{2+}$ and C$^{3+}$ as a function of radius in each model when enhanced recombination is included (middle and bottom panels of Fig. \ref{fig:cloudy}). The volume-averaged abundances differ as well, with the C$^{+}$ ionic fraction increased by a factor of 1.15--1.45 when RER is taken into account, while C$^{2+}$ decreases by factors of 1.07--1.16 and C$^{3+}$ by 1.15--1.50, depending on the model. \\

These results demonstrate that including RER in models alters the ionization balance and hence can affect elemental abundance determinations, particularly when ionization correction factors are invoked. The modifications to ionic fractions and elemental abundances could be more pronounced when RER is included for species beyond the two C ions we considered. \\ \\

\subsection{Implications of RER for radio astronomy}

Dielectronic auto-transfer represents a transition between resonant Rydberg states and doubly excited states, and thus will disturb select Rydberg populations. Collisions with neighboring Rydberg ions can return Rydberg populations to their LTE values, but our models show that dielectronic auto-transfer timescales ($\sim$ns) are much faster than Rydberg collisional timescales ($\sim$100s). \\

Rydberg emission lines from H, He, and C have been detected and used to diagnose the physical conditions of nebulae \citep{Gordon2009}. In the presence of a strong source of radio continuum emission, the lines are described by non-LTE conditions and can be affected by stimulated emission \citep{Shaver1977}. Lines emission is observed in low density ($\sim10^4$ cm$^{-3}$), low temperature ($\leq$ 10,000 K) gas, if there is a mechanism to drive the Rydberg states out of their LTE conditions \citep{Shaver1980,Anantharamaiah1993}. Dielectronic auto-transfer is a  mechanism that can cause population inversions in Rydberg states which leads to stimulated emission in their radio recombination lines \citep{Goldberg1966}. Our model for C$^{+}$ and C$^{2+}$ indicates that the Rydberg states affected by RER will produce radio emission in the range 0.112 mm to 4.41 mm and 2.92 mm to 5.14 mm respectively, but further observations are needed to verify this effect \\

\section{Conclusions}
The first empirical evidence has been presented for Rydberg Enhanced Recombination in photoionized plasmas. We predict the wavelengths and relative intensities of C II, C III, O II, and O III lines whose upper levels are populated by RER. In existing high-resolution spectra, we identify C II RER lines in the visible spectra of eight planetary nebulae, and a C III RER line in the UV spectrum of a symbiotic star. The relative strengths of the C II lines agree with predictions of our collisional-radiative models within the uncertainties.  By adjusting the recombination rate coefficients used in Cloudy to include RER contributions, we demonstrate that the impact of RER on the ionization balance and abundance determinations of astrophysical nebulae is significant. More detailed models, that account for small-scale temperature and density fluctuations, and include RER contributions for additional ions, are needed to study the extent of the implications for RER on elemental abundances. Moreover, our models indicate that RER is most prevalent at low electron temperatures, dominating other recombination mechanisms and more drastically altering the charge state distribution. Further modeling and observational efforts are needed to explore these effects.

\acknowledgments
This work was funded by NASA grant NNZ16AE97G. JG-R acknowledges support from the Severo Ochoa excellence program (SEV-2015-0548) and from the State Research Agency (AEI) of the Spanish Ministry of Science, Innovation and Universities (MCIU) and the European Regional Development Fund (FEDER) under grant AYA2017-83383-P. We thank Dr. Francis Robicheaux for his pioneering work in this area.

\bibliographystyle{aasjournal}

\end{document}